\documentclass[twocolumn]{revtex4}
\usepackage{xspace}
\usepackage{graphicx}
\usepackage{epsfig}
\usepackage{amssymb}
\begin{document}
\title{Charge transfer through single molecule contacts:\\
How reliable are rate descriptions?}
\author{D. Kast}
\email{denis.kast@uni-ulm.de} 
\author{L. Kecke}
\author{J. Ankerhold}
\affiliation{Universit\"at Ulm, Institut f\"ur Theoretische Physik, Albert-Einstein-Allee 11, 89069 Ulm, Germany}

\date{\today}

\begin{abstract}

{\bf Background: }
The trend to fabricate electrical circuits on nanoscale dimensions has led  to impressive progress in the field of
molecular electronics in the last decade. A theoretical description of molecular contacts as the building blocks of
future devices is challenging though as it has to combine properties of  Fermi liquids in the leads with charge and
phonon degrees of freedom on the molecule. Apart from ab initio schemes for specific set-ups, generic models reveal
characteristics of transport processes. Particularly appealing are descriptions based on transfer rates successfully
used in other contexts such as mesoscopic physics and intramolecular electron transfer. However, a detailed analysis
of this scheme in comparison with numerically exact data is elusive yet.

{\bf Results: }
It turns out that a formulation in terms of transfer rates provides a quantitatively accurate description even in
domains of parameter space where in a strict sense it is expected to fail, e.g.\ for lower temperatures. Typically,
intramolecular phonons are distributed according to a voltage driven steady state that can only roughly be captured
by a thermal distribution with an effective elevated temperature (heating). An extension of a master equation for the
charge-phonon complex to include effectively the impact of off-diagonal elements of the reduced density matrix
provides very accurate data even for stronger electron-phonon coupling.

{\bf Conclusion: }
Rate descriptions and master equations offer a versatile instrument to describe and understand charge transfer
processes through molecular junctions. They are computationally orders of magnitudes less expensive than elaborate
numerical simulations that, however, provide exact data as benchmarks. Adjustable parameters obtained e.g.\ from ab
initio calculations allow for the treatment of various realizations. Even though not as rigorously formulated as e.g.
nonequilibrium Greens function methods, they are conceptually simpler, more flexible for extensions, and from a practical
point of view provide accurate results as long as strong quantum correlations do not modify properties of relevant
sub-units substantially.
\end{abstract}

\keywords{Molecular contacts; inelastic charge transfer; rate equations; nonequilibrium distributions;
numerical simulations}

\maketitle

\section{Introduction}
Electrical devices on the nanoscale have received substantial interest in the last decade \cite{goser}. Impressive
progress has been achieved in contacting single molecules or molecular aggregates with normal-conducting or even
superconducting metallic leads \cite{cuniberti,scheer1}. The objective is to exploit nonlinear transport properties
of molecular junctions as the elementary units for a future molecular electronics. While the first experimental
set-ups have been operated at room temperature, meanwhile low temperatures down to the millikelvin range, the typical
regime for devices in mesoscopic solid state physics, are accessible (see e.g.\ \cite{weber1,weber2,joyez}). This
allows for detailed studies of phenomena such as inelastic charge transfer due to molecular vibrations
\cite{Park,song,webernew}, voltage driven conformational changes of the molecular backbone \cite{steigerwald},
Kondo physics \cite{franke}, and Andreev reflections \cite{joyez} to name but a few.

These developments have been accompanied by efforts to advance theoretical approaches in order to obtain an
understanding of generic physical processes on the one hand and to arrive at a tool to quantitatively describe
and predict experimental data. For this purpose, basically two strategies have been followed. One is based on
ab initio schemes that have been successfully employed for isolated molecular structures as e.g. density functional
theory (DFT). Combining DFT with nonequilibrium Greens functions (NEGF) allows to capture essential properties for
junctions with specific molecular structures and geometries \cite{datta1,datta2,cuniberti,scheer1}. This provides
insight in electronic formations along contacted  molecules and gives at least qualitatively correct results for
currents and differential conductances. However, a quantitative description on the level of accuracy known from
conventional mesoscopic devices seems out of reach yet. Further, these methods are not able to capture phenomena
due to strong correlations such as e.g.\ Kondo resonances.

Thus, an alternative route, mainly inspired by solid state methodologies, starts with simplified models that are
assumed to cover relevant physical features. The intention then is to reveal fundamental processes characteristic
for molecular electronics that give a qualitative description of observations from realistic samples, but provide
also the basis for a proper design of molecular junctions to exploit these processes. Information about specific
molecular set-ups appears merely in form of parameters which offers a large amount of flexibility. In general, to
attack the respective many body problems, perturbative schemes have  been applied, the most powerful of which are
nonequilibrium Greens functions \cite{MeirWingr,Mitra}. Conceptually simpler, easier to implement, and often better
revealing the physics are treatments in terms of master or rate equations. Being approximations to the NEGF frame
in certain ranges of parameters space, they sometimes lack the strictness of perturbation series, but have been
extensively employed for mesoscopic devices \cite{gurvitz} and quantitatively often provide data of at least
similar accuracy.
Roughly speaking, these schemes apply as long as quantum correlations between relevant sub-units of the full
compound are sufficiently weak \cite{Mitra}. Physically, it places charge transfer through molecular contacts
in the context of inelastic charge transfer through ultra-small metallic contacts (dynamical Coulomb
blockade \cite{Ingold}) and in the context of solvent or vibronic mediated intramolecular charge transfer
(Marcus theory) \cite{ankerhold1,ankerhold2,Weiss}.

While rate descriptions have been developed in a variety of formulations before
\cite{nitzan1,nitzan2,oppen,grifoni,wegewis,timm,brandes,thoss}, the performance of such a framework in comparison
with numerically exact data has not been addressed yet. The reason for that is simple: a numerical method which
provides numerically exact data in most ranges of parameters space (temperature, coupling strength, etc.) has been
successfully implemented  only very recently in form of a diagrammatic Monte Carlo approach \cite{Lothar}. Path
integral Monte Carlo methods have been used previously for intramolecular charge transfer in complex aggregates
\cite{ankerhold1,ankerhold2} in a variety of situations including correlated \cite{muehlbacher2} and externally
driven transfer \cite{muehlbacher3} and, of particular relevance for the present work, transfer in presence of
prominent phonon modes \cite{escher}.

The goal of the present work is to study a simple yet highly non-trivial set-up, namely, a molecular contact with
a single molecular level coupled to a prominent vibronic mode (phonon) which itself may or may not be embedded
in a bosonic heat bath. We develop rate descriptions of various complexity, place them into the context of NEGF,
and compare them with exact data. The essence of this study is, astonishingly enough, that rate theory provides
quantitatively accurate results for mean currents over very broad ranges of parameter space, even in domains where
they are not expected to be reliable.

\section{Results and Discussion}
In Sec.~\ref{sec1} we define the model and the basic ingredients for a perturbative treatment. A formulation
which closely follows the $P(E)$-theory for dynamical Coulomb blockade is discussed in Sec.~\ref{sec2}.
Nonequilibrium effects in the stationary phonon distribution are analyzed in Sec.~\ref{sec3} based on a dynamical
formulation of charge and phonon degrees of freedom. The presence of a secondary bath is incorporated in
Sec.~\ref{sec4} together with an improved treatment of the dot-lead coupling, which is exact for vanishing
electron-phonon interaction. The comparison with numerically exact data and a detailed discussion is given in
Sec.~\ref{sec5}.

\subsection{Model}\label{sec1}
\begin{figure}
\begin{center}
\includegraphics[width=6.5cm]{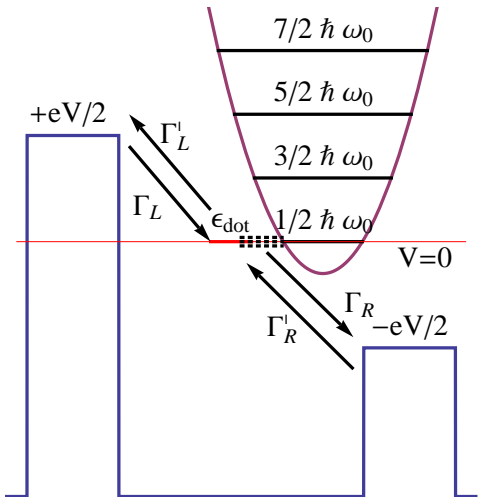}
\end{center}
\caption{Single charge transfer through a molecular contact consisting of a single electronic level coupled to a
harmonic phonon mode and contacted to metallic leads. Forward (no prime) and backward (with prime) rates are the
basic ingredients for the approximate treatment, see text for details.}
\label{Modell}
\end{figure}

We start with the minimal model of a molecular contact consisting of a single electronic level coupled to fermionic
reservoirs, where a prominent internal molecular phonon mode interacting with the excess charge is described by a
harmonic degree of freedom (cf.~Fig.~\ref{Modell}) \cite{Flensberg,Mitra,Egger}.
Neglecting spin degrees of freedom the total compound is thus described by
\begin{eqnarray}
H & = & H_{L/R}+H_T+H_D+H_{D,Ph}+H_{Ph} \nonumber \\
& = & \sum_{\alpha=L, R;\ k} \epsilon_{k, \alpha} c_{k, \alpha}^{\dagger}c_{k, \alpha}
+\sum_{\alpha=L, R;\ k}(T_{k, \alpha} c_{k, \alpha}^{\dagger} d + h.c.) \nonumber \\
&& + {\epsilon}_D d^{\dagger} d  + \frac{p_0^2}{2m}+\frac{m\omega_0^2}{2} \left(x_0+ l_0 d^\dagger d\right)^2
\label{Hamiltonian}
\end{eqnarray}
Here, the $T_{k, \alpha}$ denote tunnel couplings between dot level and reservoir $\alpha$ and $l_0=M_0
\sqrt{2/\hbar\omega_0^3 m}$ contains the coupling $M_0$ between excess charge and phonon mode. An external
voltage $V$ across the contact is applied symmetrically around the Fermi level such that
$\epsilon_{k, \alpha}=\epsilon_0(k)+\mu_\alpha$ with the bare electronic dispersion relation $\epsilon_0(k)$
and chemical potentials $\mu_L=+eV/2, \mu_R=-eV/2$.
Below, this model will be further extended  to include the embedding of the prominent mode into a large reservoir
of residual molecular and/or solvent degrees of freedom acting as a heat bath. Qualitatively, since the dot
occupation $d^\dagger d$ can only take the values $q=0, 1$, the sub-unit $H_D+H_{D,Ph}+H_{Ph}$ describes a two
state system coupled to a harmonic mode (spin-boson model \cite{Weiss}). Depending on the charge state of the
dot the phonon mode is subject to potentials $V_q(x_0)=(m\omega_0^2/2)(x_0+l_0 q)^2$.
Now, the presence of the leads  acts (for finite voltages) as an external driving force to alternately
charging ($q=1$) and discharging ($q=0$) the dot, thus switching alternately between $V_0$ and $V_1$ for the
phonon mode. The classical energy needed to reorganize the phonon is the so-called reorganization energy
$\Lambda=V_1(0)-V_0(0)=M_0^2/\hbar\omega_0$. Quantum mechanically, the phonon mode may also tunnel through
the energy barrier located around $x_0=-l_0/2$ separating the minima of $V_{0, 1}$.

It is convenient to work with dressed electronic states on the dot and thus to apply a polaron transformation
generating the shift $l_0$ in the oscillator coordinate associated with a charge transfer process, i.e.,
\begin{eqnarray}
U=\exp\left(-i \frac{p_{0}l_0}{\hbar}d^{\dagger}d \right) \label{Polaron}
\end{eqnarray}
with momentum operator $p_0=i\sqrt{\hbar m\omega_0/2}(b_{0}^{\dagger} - b_{0})$ where $b_0^\dagger, b_0$ are
creation and annihilation operators of the phonon mode, respectively. We mention in passing that complementary
to the situation here, the theory of dynamical Coulomb blockade in ultra-small metallic contacts is based on a
transformation which generates a shift in momentum (charge) rather than position \cite{Ingold}. Now, the
electron-phonon interaction is completely absorbed in the tunnel part of the Hamiltonian, thus capturing
 the cooperative effect of charge tunneling onto the dot and photon excitation in the molecule,
 i.e.\ $\tilde{H}=H_{L/R}+{H}_D+\tilde{H}_{Ph}+\tilde{H}_T$ with
\begin{eqnarray}
\label{HPMTpol}
\tilde{H}_{Ph}&=&\hbar\omega_{0} \left(b_{0}^{\dagger} b_{0}+\frac{1}{2}\right)\nonumber\\
\tilde{H}_T & =&  \sum_{\alpha=L, R;\ k} T_{k, \alpha} c_{k, \alpha}^{\dagger}d \;
\exp\left(\frac{i}{\hbar} p_0 l_0 \right)+h.c.\,
\end{eqnarray}

Single charge tunneling through the device can be captured formally exactly under weak conditions (e.g.\ instantaneous
equilibration in the leads during charge transfer) within the Meir-Wingreen formulation based on nonequilibrium
Greens functions \cite{MeirWingr,Mitra}. For the current voltage characteristics one finds
\begin{eqnarray}
I(V) & = & \frac{4e}{\hbar}{\displaystyle \int d\epsilon \frac{\Sigma_L\Sigma_R}{\Sigma_L + \Sigma_R} \left[f_{\beta}\left(\epsilon-\frac{eV}{2}\right)-f_{\beta}\left(\epsilon+\frac{eV}{2}\right)\right]} \nonumber \\
&& \;\;\;\;\;\;\; \times\left[ i G^>(\epsilon)-iG^<(\epsilon)\right]
\end{eqnarray}
with energy dependent lead self-energies $\Sigma_{\alpha}(\epsilon)=2\pi\sum_{k}
|T_{k, \alpha}|^2\delta(\epsilon-\epsilon_k)$ and with the Fourier transforms of the time dependent Greens
functions $G^<(t)  =  i\langle d^{\dagger}d(t)\rangle$ and $G^>(t)  = -i\langle d(t)d^{\dagger} \rangle$.
Upon applying the polaron transformation (\ref{Polaron}), one has
\begin{eqnarray}
  \tilde{G}^<(t) &=& i\left\langle d^{\dagger} e^{-\frac{i}{\hbar}p_0 l_0}  e^{\frac{i}{\hbar}p_0(t) l_0}d(t)
\right\rangle \nonumber  \\
  \tilde{G}^>(t) &=& -i\left\langle e^{\frac{i}{\hbar}p_0(t) l_0} d(t)d^{\dagger}
e^{-\frac{i}{\hbar}p_0 l_0} \right\rangle\, ,
\end{eqnarray}
where all expectation values are calculated with the full Hamiltonian (\ref{HPMTpol}). Of course, for
$T_{k, \alpha}\to 0$, the Greens functions factorize such as e.g.\  $\tilde{G}^<(t) \to i\langle d^{\dagger}
d(t)\rangle_D \ \exp[J(t)]$ with the phonon correlation
 \begin{equation}
 \label{phononcorr}
 {\rm e}^{J(t)}= \left\langle{\rm e}^{-\frac{i}{\hbar}p_0 l_0}
 {\rm e}^{\frac{i}{\hbar}p_0(t) l_0}\right\rangle_{Ph}
  \end{equation}
into expectation values with respect to the dot (D) and the phonon (Ph), respectively. Any finite tunnel coupling
induces correlations that in analytical treatments can only be incorporated perturbatively. There, the proper
approximative scheme depends on the range of parameter space one considers. Generally speaking, there are four
relevant energy scales  $\Sigma_{L/R}$, $M_0$, $k_{\rm B} T$, and $\hbar\omega_0$ of the problem corresponding
to three independent dimensionless parameters, e.g.,
\begin{equation}
m_0=\frac{M_0}{\hbar\omega_0}\ , \ \theta=\omega_0\hbar\beta\ , \ \sigma=\frac{\Sigma_L+\Sigma_R}{\hbar\omega_0}\, .
\end{equation}
In the sequel we are interested in the low temperature domain $\theta>1$ where thermal broadening of phonon
levels is small so that discrete steps appear in the $IV$-characteristics. Qualitatively, seen from the dynamics
of the phonon mode, two regimes can be distinguished according to the adiabaticity parameter
$\Sigma/\hbar\omega_0=\sigma$: For $\sigma<1$ the phonon wave packet fulfills on a {\em given surface} $V_0$ or $V_1$ multiples of oscillations  before a charge transfer process happens to occur. The electron carries excess energy due to  a finite voltage
which may be absorbed by the phonon to reorganize to the new conformation (in the classical case the
reorganization energy $\Lambda$). In the language of intramolecular charge transfer this scenario corresponds
to the diabatic regime with well-defined surfaces $V_q$. In the opposite regime $\sigma>1$ charge transfer
is fast so that the phonon may obey multiples of switchings {\em between the surfaces} $V_{0, 1}$ .
This is the adiabatic regime. In this latter range the impact of the adiabaticity on
the diabatic ground state wave functions is weak for $m_0<1$ when the distance of the diabatic surfaces is small
compared to the widths of the ground states. For $m_0>1$ in both regimes electron transfer is accompanied by
phonon tunneling through energy barriers separating minima of adiabatic or diabatic surfaces. The dynamics of
the total compound is then determined by  voltage driven collective tunneling processes.
Master equation approaches to be investigated below, rely on the assumption that both sub-units, charge degree
of freedom and phonon mode, basically preserve their bare physical properties even in case of finite coupling
$m_0$. Hence, since the model (\ref{Hamiltonian}) can be solved exactly in the limits $m_0=0$ and $\sigma=0$
and following the above discussion, we expect them to capture the essential physics quantitatively in the domain
$m_0<1$ and for all ratios $\sigma$. We note that recently the strong coupling limit including the current statistics has been addressed as well \cite{komnik1,komnik2}.

\subsection{Rate approach I}\label{sec2}

The simplest perturbative approach considers the cooperative effect of electron tunneling and phonon excitation
in terms of Fermi's golden rule for the tunneling part $\tilde{H}_T$. For this purpose one derives transition
rates for sequential transfer according to Fig. \ref{Modell}. A straightforward calculation for energy independent
self-energies $\Sigma_{L/R}$ (wide-band limit) gives the forward rate onto the dot from the left lead
\begin{eqnarray}
  \Gamma_L(V,\epsilon_D)=\frac{\Sigma_L}{\hbar}{\displaystyle \int d\epsilon f_{\beta}\left(\epsilon-\frac{eV}{2}\right)}
  P_0(\epsilon-\epsilon_D)\, ,
  \label{Gamma1_allg}
\end{eqnarray}
where  $f_\beta(\epsilon)$ is the Fermi distribution. Inelastic tunneling associated with energy emission /
absorption of phonons is captured by the Fourier transform of the phonon-phonon correlation $\exp[J(t)]$ leading to
\begin{equation}
  P_0(\epsilon)  =  e^{-\rho_{a}-\rho_{e}}{\displaystyle \sum_{k,l}\frac{\rho_{a}^k \rho_{e}^l}{k! l!}\delta
  \left[\epsilon-\hbar\omega_0(l-k)\right] \label{P(E)}}
\end{equation}
with $\rho_{a/e}=(m_0^2/2)\left[\coth\left(\frac{\theta}{2}\right)\mp 1\right]$ denoting mean values for single
phonon absorption (a) and emission (e). The exponentials in the prefactor contain the dimensionless reorganization
energy $m_0^2=\Lambda/\hbar\omega_0$.
Apparently, inelastic charge transfer includes the exchange of  multiple phonon quanta according to a Poissonian
distribution. Further, one has the detailed balance relation $P_0(-\epsilon)={\rm e}^{-\beta \epsilon}
P_0(\epsilon)$.
For vanishing phonon-electron coupling $m_0\to 0$ only the elastic peak survives $P_0(\epsilon)\to \delta(\epsilon)$.
We note again the close analogy to the $P(E)$-theory for dynamical Coulomb blockade \cite{Ingold}. Moreover, golden
rule rates for intramolecular electron transfer between donor and acceptor sites coupled to a single phonon mode
are of the same structure with the notable difference, of course, that in this case one has discrete density of
states for both sites \cite{Weiss,nitzan2}. The fundamental assumption underlying the golden rule treatment is
that equilibration of the phonon mode occurs much faster than charge transfer. In the last two situations this
is typically guaranteed by the presence of a macroscopic heat bath (secondary bath) strongly coupled to the
prominent phonon mode. Here, the fermionic reservoirs in the leads impose phonon relaxation only due to charge
transfer. Thus, for finite voltage the steady state is always a nonequilibrium state that can only roughly be
described by a thermal distribution of the bare phonon system (see below). One way to remedy this problem is to
introduce a phonon-secondary bath interaction as well, see below in Sec.~\ref{sec4}. The remaining transition
rates easily follow due to symmetry
\begin{eqnarray}
  \Gamma_R(V,\epsilon_D) & = & \frac{\Sigma_R}{\Sigma_L}\, \Gamma_L(V,-\epsilon_D)\ , \nonumber \\
  \Gamma'_R(V,\epsilon_D) & = & \frac{\Sigma_R}{\Sigma_L}\, \Gamma_L(-V,\epsilon_D)\ ,\nonumber \\
  \Gamma'_L(V,\epsilon_D) & = & \Gamma_L(-V,-\epsilon_D)\, .
\end{eqnarray}

\begin{figure}
\begin{center}
\includegraphics[width=8cm]{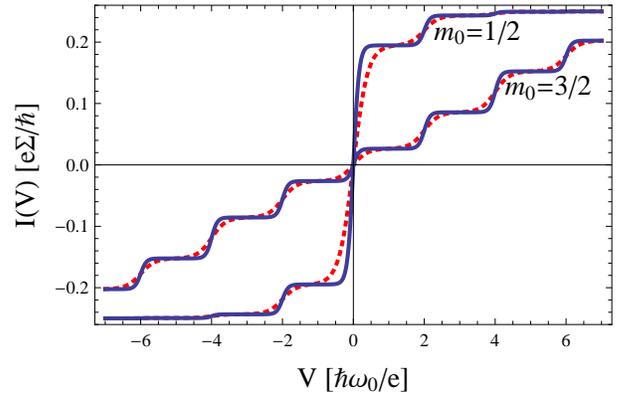}
\end{center}
\caption{$IV$-characteristics for symmetric coupling $\Sigma_L=\Sigma_R$ and for varying electron-phonon
coupling $m_0$ at inverse temperature $\theta=25$ (solid) and $\theta=10$ (dashed).}
\label{I(V)-Kurven}
\end{figure}

Now, summing up forward and backward events, the dot population follows from
\begin{equation}
\label{masterdot}
\frac{d\hat{p}_{\rm dot}}{dt}=-\Gamma_{\rm tot, 0}\, \hat{p}_{\rm dot}+\Gamma_{\rm d}
\end{equation}
with the total rate $\Gamma_{\rm tot, 0}=\Gamma_L+\Gamma_R+\Gamma_L'+\Gamma_R'$  and the rate for transfer
towards the dot $\Gamma_{\rm d}=\Gamma_L+\Gamma_R'$ obtained according to (\ref{Gamma1_allg}). Note that
for vanishing electron-phonon coupling $M_0=0$ one has $\hbar\Gamma_{\rm tot, 0}(M_0=0)=\Sigma_L+\Sigma_R$.
The steady state distribution $\hat{p}_{\rm dot}\to p_{\rm dot}=\Gamma_{\rm d}/\Gamma_{\rm tot, 0}$ is
approached with relaxation rate $\Gamma_{\rm tot, 0}$. For a symmetric situation $\Sigma_L=\Sigma_R$ with
$\epsilon_D=0$ one shows that $p_{\rm dot}=1/2$ independent of the voltage, while asymmetric cases lead to
voltage dependent stationary populations.  The steady state current is given by
$I(V)=(e/2) [(\Gamma_L-\Gamma_R') (1-p_{\rm dot}) - (\Gamma_L'-\Gamma_R) p_{\rm dot}]$ so that
\begin{equation}
I(V)=e \frac{\Gamma_L \Gamma_R - \Gamma_L' \Gamma_R'}{\Gamma_{\rm tot, 0}}\, .
\end{equation}
A transparent expression is obtained for $\epsilon_D=0$, namely,
\begin{eqnarray}
I(V) & = & \frac{e}{\hbar}\frac{\Sigma_L \Sigma_R}{\Sigma_L+\Sigma_R} \times \nonumber \\
&& {\displaystyle \int d\epsilon 
\left[f_{\beta}\left(\epsilon-\frac{eV}{2}\right)-f_{\beta}\left(\epsilon+\frac{eV}{2}\right) \right] P_0(\epsilon). \, \nonumber} \\
  \label{StromGammaLR}
\end{eqnarray}
Despite its deficiencies mentioned above, the golden rule treatment provides already a qualitative insight in the
transport characteristics. Typical results are shown in Fig. \ref{I(V)-Kurven}.
The $IV$-curves display the expected steps at $eV=2n\hbar\omega_0, n\in {\cal Z}$. Each time the voltage
$\frac{eV}{2}$ exceeds multiples of $\hbar\omega_0$ new transport channels open associated with the
excitation of one additional phonon. For higher temperatures the steps are smeared out by thermal fluctuations.
The range of validity of this description follows from the fact that a factorizing assumption for the
phonon-electron correlation has been used {\em and} an instantaneous equilibration of the phonon mode after a
charge transfer, that means $\sigma<1$ and  $m_0<1$. The latter constraint guarantees that conformational
changes of the phonon distribution remain small.

There are now three ways to go beyond this golden rule approximation. With respect to the phonon mode, one
is to explicitly account for its nonequilibrium dynamics, another is to introduce a direct interaction
with a secondary heat bath in order to induce sufficiently fast equilibration. With respect to the dot
degree of freedom one can exploit the fact that for vanishing charge-phonon coupling the model can be solved
exactly.

\subsection{Master equation for nonequilibrated phonons}\label{sec3}

To derive an equation of motion for the combined dynamics of charge and phonon degrees of freedom, one starts
from a Liouville-von Neumann equation for the full polaron transformed compound (\ref{HPMTpol}). Then,
applying a Born-Markov type of approximation with respect to the tunnel coupling to the fermionic reservoirs,
one arrives at a Redfield-type of equation for the reduced density matrix of the dot-phonon system \cite{Mitra}.
An additional rotating wave approximation (RWA) separates the dynamics of diagonal (populations) and
off-diagonal (coherences) elements of the reduced density. Denoting with $\hat{P}_q^n$ the probability to find
$q$ charges on the dot (here, for single charge transfer $q=0, 1$) and the phonon in its $n$-th eigenstate,
one has
\begin{eqnarray}
  \frac{d\hat{P}_q^n(t)}{dt} & = &
  -\frac{1}{\hbar} \sum_{\alpha=L, R;\ k} |f_{n,k}|^2\Sigma_\alpha \times \nonumber \\ 
 && \;\; \left[f_{\beta}\left(E_{kn}^{q, \alpha}\right) \hat{P}_q^n-f_{\beta}
  \left( -E_{kn}^{q, \alpha}\right)\hat{P}_{q+\nu_q}^{k}\right]
  \label{NEQMaster1}
\end{eqnarray}
with $\nu_0=1, \nu_1=-1$ and  energies $E_{kn}^{q, \alpha}=\hbar\omega_0(k-n)+\nu_q (\mu_\alpha +\epsilon_D)$.
The matrix elements of the phonon shift operator $f_{n,k}=\langle n|\exp(i p_0 l_0/\hbar) |k \rangle$ read
\begin{eqnarray}
  f_{n,k} & = &  {\rm e}^{-m_0^2/2}\frac{(-m_0)^{|n-k|}}{(n-k)!} 
  \left(\prod_{l=\min(n,k+1)}^{\max(n,k)} l \right)^{\frac{1}{2}} \times \nonumber \\
&& {}_1 F_1 \left(\ max(n,k)+1,|n-k|+1,-m_0^2 \right)\, ,
\end{eqnarray}
where ${}_1 F_1$ denotes a hypergeometric function. The underlying assumptions of this formulation require weak
dot-lead coupling $\sigma<1$ and sufficiently elevated temperatures $\sigma\theta<1$ for a Markov approximation
to be valid. We will see below when comparing low temperature results with numerically exact data that this seems
to be only a weak constraint though.

\begin{figure}
\begin{center}
\includegraphics[width=8cm]{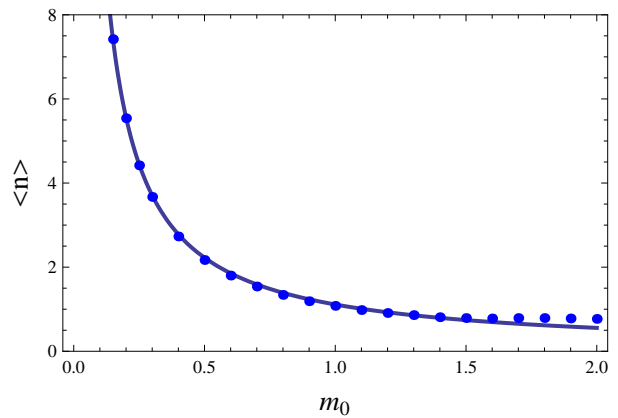}
\end{center}
\caption{Mean phonon number in nonequilibrium for $eV=3\hbar\omega_0$ and vs. the electron-phonon coupling $m_0$.}
\label{Erwartungswert}
\end{figure}

The calculation of the steady state distribution $P_q^n=\lim_{t\to\infty} \hat{P}_q^n(t)$ reduces to a standard
matrix inversion. One can show that for a symmetric system with $\epsilon_D=0,\ \Sigma_L=\Sigma_R$ one has
$P_0^n=P_1^n$. A typical example for the mean phonon number $\langle n\rangle=\sum_{q,n} n P_q^n$ is
depicted in Fig. \ref{Erwartungswert}. The curve is well approximated by $a/m_0$ with $a\approx 0.7$.
Apparently, $\langle n\rangle$ diverges for $m_0\to 0$ since then $P_0^n, P_1^n$ approach constants independent
of the phonon number.
Upon closer inspection one finds that excitation is more likely than absorption, i.e\ $f(n,n+1)>f(n,n-1)$,
for all $0\leq n\leq N_0(m_0)$ where $N_0(m_0)$ increases with decreasing $m_0$. The opposite is true for
$n>N_0(m_0)$ so that in a steady state, depending on the voltage, the tendency is to have higher excited phonon
states occupied for smaller couplings $m_0$. In particular, for strong coupling transitions $n\to n+k, k\geq 0$
are basically blocked at quite small $n$.

\begin{figure}
\begin{center}
\includegraphics[width=8cm]{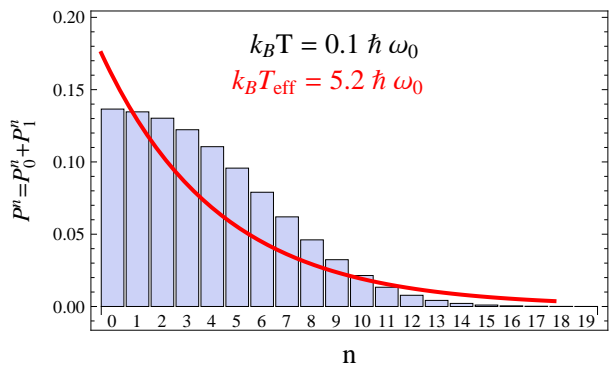}
\end{center}
\caption{Phonon number distribution in nonequilibrium  for $eV=5\hbar\omega_0$, $m_0=0.5$ and 
$k_{\rm B} T/\hbar\omega_0=0.1$ (histogram). The solid line depicts a fit to a Boltzmann distribution. See text
for details.}
\label{Boltzmann}
\end{figure}

A convenient strategy to include nonequilibrium effects in the phonon distribution, sometimes used in the
interpretation of experimental data, is the introduction of an effective temperature $T_{\rm eff}$. This way,
one could return to the simpler modeling of the previous section. However, the procedure to identify
$P_0^n+P_1^n \approx P_\beta^n=\exp(-\beta_{\rm eff}\hbar\omega_0 n)/[1-\exp(-\beta_{\rm eff}\hbar\omega_0)]$
is not reliable as Fig. \ref{Boltzmann} reveals. While it clearly shows the general tendency of a substantial
heating of the phonon degree of freedom induced by the electron transfer, the profile of a thermal distribution
strongly differs from the actual steady state distribution.

\begin{figure}
  \begin{center}
    \includegraphics[width=8cm]{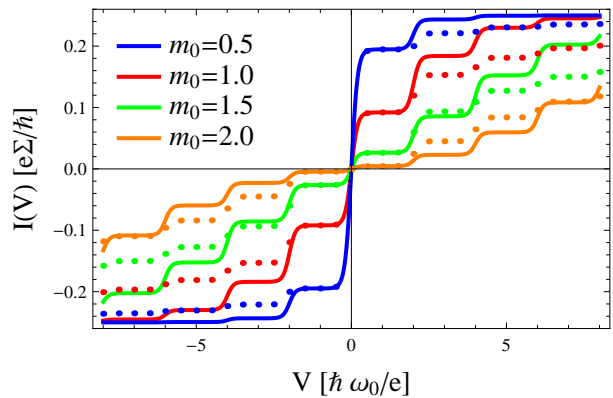}
  \end{center}
  \caption{$IV$-characteristics for equilibrated (solid) and nonequilibrated (dotted) phonon distributions
   according to (\ref{StromGammaLR}) and (\ref{mastercurrent}), respectively.}
  \label{Strom-VGL}
\end{figure}

Non-equilibrated phonons leave their signatures also in the $IV$-curves as compared to equilibrated ones.
The net current through the contact follows from summing up the transfer rates from / onto the dot according
(\ref{NEQMaster1}), namely,
\begin{eqnarray}
\label{mastercurrent}
I & = & \frac{e}{2\hbar} \sum_{n, k} |f_{n, k}|^2 \left\{\left[\Sigma_L f_\beta\left(E_{nk}^{0, L}\right) -\Sigma_R f_\beta\left(E_{nk}^{0, R}\right)\right] P_0^n \right. \nonumber \\
&& \;\;\;\; - \left. \left[\Sigma_L f_\beta\left(E_{nk}^{1, L}\right) -\Sigma_R f_\beta\left(E_{nk}^{1, R}\right)\right] P_1^n\right\}\, .
\end{eqnarray}
Fig. \ref{Strom-VGL} shows that deviations are negligible for low voltages in the regime around the first resonant
step ($|eV/2|<\hbar\omega_0$), where at sufficiently low temperatures only the ground state participates so that
the steady state distribution basically coincides with the thermal one. For larger voltages deviations occur with
the tendency that for smaller couplings $m_0$ the nonequilibrated current is always smaller than the equilibrated
one ($I_{\rm non}< I_{\rm eq}$), while the opposite scenario ($I_{\rm non}> I_{\rm eq}$) is observed for larger
$m_0$. At sufficiently large voltages, one always has $I_{\rm non}<I_{\rm eq}$.
This behavior results from the combination of two ingredients, namely, the phonon distributions $P_q^n$ and the
Frank-Condon overlaps $|f_{n, k}|^2$. To see this in detail, let us consider a fixed voltage. Then, on the one
hand, for smaller $m_0$ the steady state distribution is broad (cf.~Fig. \ref{Erwartungswert}) so that due to
normalization less weight is carried by lower lying states compared to a thermal distribution at low temperatures;
on the other hand, for $m_0<1$ the overlaps $|f_{n, k}|^2$ favor contributions from low lying states in the
current (\ref{mastercurrent}) which is thus smaller than $I_{\rm eq}$. For increasing electron-phonon coupling
$m_0>1$ overlaps $|f_{n, k}|^2$ tend to include broader ranges of phonon states also covered by $P_q^n$
compared to those of low temperature thermal states. A voltage dependence arises since with increasing voltage
higher lying phonon states participate in the dynamics supporting the scenario for smaller couplings.
Interestingly, as already noted in \cite{oppen} the overlaps $|f_{n, k}|^2$ may vanish for certain combinations
of $n ,m$ depending on $m_0$ due to interferences of  phonon eigenfunctions localized on different diabatic
surfaces $V_q, q=0, 1$.

\subsection{Rate approach II}\label{sec4}

The assumption of a thermally distributed phonon degree of freedom during the transport can be physically
justified only if this mode interacts directly and sufficiently strongly with an additional heat bath
(secondary bath) realized e.g. by residual molecular modes. Here we will generalize the formulation of
Sec.~\ref{sec2}  to a situation where the secondary bath is characterized by Gaussian fluctuations.
Its corresponding modes can thus effectively be represented by a quasi-continuum of harmonic oscillators
for which the phonon correlation function (\ref{phononcorr}) can be calculated easily
\begin{equation}
J(t)= {\displaystyle \int_0^\infty \frac{d\omega}{\pi} \frac{I(\omega)}{\omega^2}\,
\left[{\rm coth}\left(\frac{\omega\hbar\beta}{2}\right)[\cos(\omega t)-1]-i \sin(\omega t)\right]\,} .
\end{equation}
Here the spectral density $I(\omega)$ describes now the combined distribution of the prominent mode and its secondary bath.
It is thus proportional to the imaginary part of the dynamical susceptibility of a damped harmonic oscillator \cite{Weiss}.
For a purely ohmic distribution of bath modes, one has
\begin{equation}
  I(\omega) =  2m_0^2\omega_0^3\, \frac{\gamma\omega}{(\omega^2-\omega_0^2)^2 + \gamma^2 \omega^2}\, ,
\end{equation}
where $\gamma$ denotes the coupling between phonon mode and bath. The Fourier transform of $\exp(J)$ reads at finite temperatures
\begin{eqnarray}
  P_{\gamma}(\epsilon) & = & {\rm e}^{-\rho_{\gamma}} \delta(\epsilon) + \frac{e^{-\rho_{\gamma}}}{\pi}\times  \nonumber \\
  &&  \Re {\displaystyle \sum_{(k,l)\not= (0,0)}\frac{\rho_{\gamma,a}^k}{k!}\frac{\rho_{\gamma,e}^l}{l!} 
  \frac{i}{\epsilon+\hbar\Omega k-\hbar\Omega^* l}}
  \label{Lorentz}
\end{eqnarray}
with frequency $ \Omega = \omega_0 \xi+i \gamma/2$ and  
$\rho_{\gamma,a}(\Omega)= (m_0^2/2\xi\,\Omega^2)[\coth(\beta\hbar\Omega/2)-1]$ where 
$\xi=\sqrt{1-\gamma^2/4\omega_0^2}$. Further, $\rho_{\gamma,e}(\Omega)=-\rho_{\gamma,a}(-\Omega^*)$ 
(* means complex conjugation) and $\rho_{\gamma}=\Re[\rho_{\gamma,a}+\rho_{\gamma,e}]/2$.
In the above expression contributions from the Matsubara frequencies in Equation (17) have been neglected since they are only relevant in the regime $\gamma \hbar \beta \gg 2\pi$ which is not studied here.
Apparently, the coupling to the bosonic bath effectively induces a broadening of the dot levels
$\hbar\gamma (k+l)/2$ compared to the purely elastic case (\ref{P(E)}). In the low temperature regime, where
for equilibrated phonons absorption (related to $k$) is negligible, the widths grow proportional to $l$. The
presence of the secondary bath drives the prominent phonon mode towards thermal equilibrium with a rate
proportional to this broadening. Hence, if the time scale for thermal relaxation is sufficiently smaller than
the time scale for charge transfer, i.e.\  $1/\tau_l\equiv\Sigma_L+\Sigma_R/\gamma\ll 1$, the assumption of
an equilibrated phonon mode is justified and the golden rule formulation (\ref{StromGammaLR}) can be used with
$P_0(\epsilon)\to P_\gamma(\epsilon)$. However, this argument no longer applies in the overdamped situation
$\gamma/\omega_0\gg 1$, where the phonon mode exhibits a sluggish thermalization on the time scale
$\gamma/\omega_0^2$ which may easily exceed $\tau_l$.

As already mentioned above, for vanishing charge-phonon coupling $m_0=0$, the model (\ref{HPMTpol}) can be
solved exactly to all orders in the lead-dot coupling \cite{Mitra}. In the frame of  a rate description,
one observes that in this limit the dot population (\ref{masterdot}) decays proportional to $\Sigma_L+\Sigma_R$.
The golden rule version of the theory  neglects this broadening in (\ref{StromGammaLR}) since it is associated
with higher order contributions to the current (\ref{StromGammaLR}). Now, recalling that $P_0(\epsilon)$
reduces to  a $\delta$-function for $m_0\to 0$, this finite lifetime of the electronic dot level is included
to {\em all orders} by performing the time integral in the Fourier transform with
$\epsilon\to \epsilon - i (\Sigma_L+\Sigma_R)/2\equiv \epsilon- i \Gamma_{\rm tot}(M_0=0)/2$
[see (\ref{masterdot})]. In fact, this way one reproduces the {\em exact} solution  (one electronic level
coupled to leads with energy independent couplings), i.e., its exact spectral function.  To be specific, let
us restrict ourselves in the remainder to the symmetric situation $\Sigma_L=\Sigma_R\equiv \Sigma/2,$ and
$ \epsilon_D=0$. Then, in the presence of the phonon mode ($m_0\neq 0$) the corresponding function
$P_0^\Sigma(\epsilon)$ follows from (\ref{P(E)}) by replacing the $\delta$-function by
$i/[\epsilon+\hbar\omega (k-l)+i \Sigma/2]$. Again following the spirit of a rate treatment, an improved version
of this result accounting for higher order electron-phonon correlations is obtained by using instead of the
bare dot level width $\Sigma/\hbar\equiv\Gamma_{\rm tot, 0}(M_0=0)$, the decay rate
$\Gamma_{\rm tot, 0}(M_0\neq 0)$. Equivalently, one replaces
$i/[\epsilon+\hbar\omega (k-l)+i \Sigma/2]\to i/[\epsilon+\hbar\omega (k-l)+i\hbar\Gamma_{\rm tot, 0}/2]$
to arrive at an improved $P_1^\Sigma(\epsilon)$. We note that within a Greens-function approach and upon
approximating the corresponding equations of motion a similar result has been found in \cite{Mitra,Flensberg}
with the difference though that there instead of $\Gamma_{\rm tot, 0}$ an imaginary part of a phonon state
dependent self-energy $\Sigma''_{k,l}$ appears. One can show that the  $\Gamma_{\rm tot, 0}$ appearing
here within a rate scheme is related to a thermally averaged $\Sigma''_{k,l}$.

Now, an additional secondary bath can be introduced as above by combining (\ref{Lorentz}) with $P_1^\Sigma$,
leading eventually to
 \begin{equation}
  P_{\gamma, 1}^\Sigma(\epsilon)  =  \frac{e^{-\rho_{\gamma}}}{\pi\hbar} \Re {\displaystyle \sum_{(k,l)
  \geq (0,0)}\frac{\rho_{\gamma,a}^k}{k!}\frac{\rho_{\gamma,e}^l}{l!} 
  \frac{i}{\frac{\epsilon}{\hbar}+\Omega k-\Omega^* l+i\frac{\Gamma_{\rm tot, 0}}{2}}}.
  \label{P(E)final}
\end{equation}
The width of the electronic dot level is thus voltage dependent and approaches the bare width from below for
large voltages, that is
 $\lim_{V\to\infty} \Gamma_{\rm tot, 0}(V)=\Sigma/\hbar$.
The range of validity of this scheme is the following: it applies to all ratios $\sigma/m_0$ in the domain
where the electron-phonon coupling is weak $m_0<1$. For $m_0>1$ charge transfer is strongly suppressed and
the phonon dynamics still occurs on diabatic surfaces for $\sigma/m_0\ll 1$ so that we expect the approach
to cover this range as well.

\subsection{Comparison with numerically exact results}\label{sec5}

A numerically exact treatment of the nonequilibrium dynamics of the model considered here is a formidable task.
The number of formulations which allow simulations in non-perturbative ranges of parameter space is very limited.
Among them is a recently developed diagrammatic Monte Carlo approach (diagMC) based on a numerical evaluation of
the full Dyson series which in contrast to numerical renormalization group (NRG) methods \cite{bulla}  covers
the full temperature range. For the sector of single charge transfer results have been obtained with and without
the presence of a secondary bath interacting with the dot phonon mode. We note that computationally these
simulations are very demanding as for each parameter set and a given voltage the stationary current for the
$IV$ curve needs to be extracted from the saturated value of the time dependent current $I(t)$ for longer
times. Typical simulation times are on the order of several days and even up to weeks depending on the parameter
range. In contrast, rate treatments require minimal computational efforts and can be done within minutes.
Here, we compare numerically exact findings with those gained from the various types of rate/master equations
discussed above.

\begin{figure}
\begin{center}
\includegraphics[width=8.5cm]{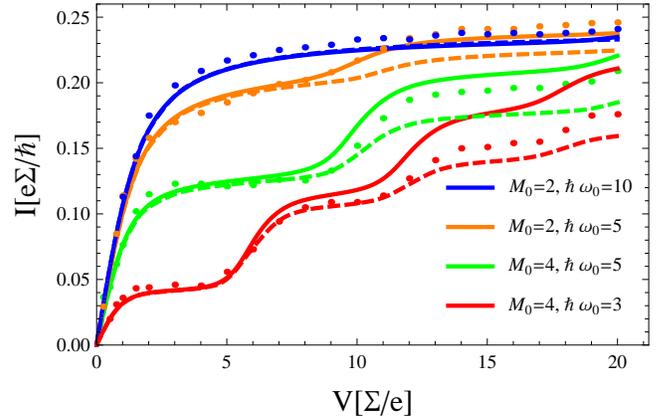}
\end{center}
\caption{$IV$-characteristics according to approximate models based on equilibrated phonons (solid) and
nonequilibrated phonons (dashed) together with exact DQMC data (dots) for $k_{\rm B} T/\Sigma=0.2$ and
without coupling to a secondary bath ($\gamma=0$). All quantities are scaled with respect to the dot-lead
coupling $\Sigma$.}
\label{MC2}
\end{figure}
We start with the scenario where the coupling to a secondary bath is dropped ($\gamma=0$) to reveal the
impact of nonequilibrium effects in the phonon mode. The formulation for an equilibrated phonon is based
on (\ref{StromGammaLR}) with $P_0$ replaced by $P_{\gamma, 1}^\Sigma$ in (\ref{P(E)final}), while the
steady state phonon distribution is obtained from the stationary solutions to (\ref{NEQMaster1}). In the
latter approach the intrinsic broadening of the dot electronic level due the lead coupling is introduced
in the following way:
One first determines via (\ref{NEQMaster1}) a steady state distribution $P_q^n$. This result is used
for an effective self-energy contribution (total decay rate) for non-equilibrated phonons, i.e.,
\begin{equation}
  \Gamma_{\rm tot, neq}(V) = \frac{\Sigma}{\hbar}{\displaystyle \sum_{\alpha=L, R;\ n,k,q}|f_{n,k}|^2 
  P_q^n f_{\beta}\left(E_{nk}^\alpha\right)}\, ,
\end{equation}
where $E_{nk}^\alpha=\hbar\omega_0 (n-k)+\mu_\alpha$. We note in passing that $\lim_{V\to \infty}
\Gamma_{\rm tot, neq}(V)=\Sigma/\hbar\equiv \Gamma_{\rm tot}(M_0=0)$. Subsequently, an improved result for
the steady state phonon distribution at a given voltage is evaluated working again with (\ref{NEQMaster1})
but using the replacement
\begin{eqnarray}
\label{replacement}
  &&f_{\beta}\left(\hbar\omega_0(k-l) \mp\frac{eV}{2}\right) \to \nonumber \\
 &&{\displaystyle \int \frac{d\epsilon}{2\pi} f_{\beta}\left(\epsilon\mp\frac{eV}{2}\right) \nonumber
\frac{\hbar\Gamma_{\rm tot, neq}}{[\epsilon-\hbar\omega_0(k-l)]^2 + \hbar^2\Gamma_{\rm tot, neq}^2/4}}\,. \\
\end{eqnarray}
Of course, for $\Sigma\to 0$ the standard Fermi distribution is regained.
The corresponding steady state phonon distribution eventually provides the current according to
(\ref{mastercurrent}) using in this expression the same replacement (\ref{replacement}). The procedure
relies on weak electron-phonon coupling $m_0<1$ and requires in principle also sufficiently elevated temperatures.

\begin{figure}
\begin{center}
\includegraphics[width=8.5cm]{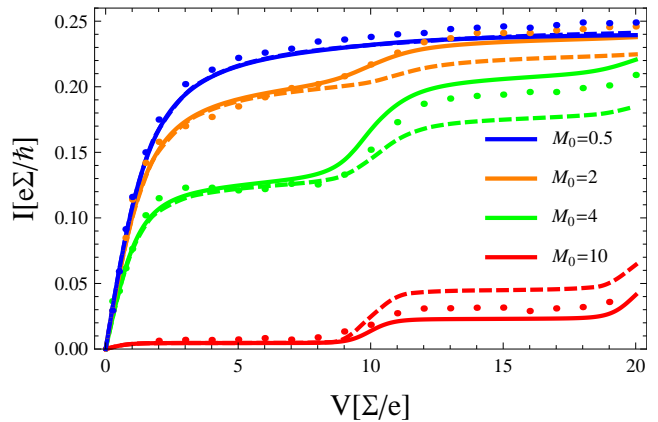}
\end{center}
\caption{Same as in Fig.~\ref{MC2} but for fixed $\hbar\omega_0/\Sigma=5$ and varying electron-phonon coupling.}
\label{MC3}
\end{figure}

Results are shown in Fig.~\ref{MC2} together with corresponding diagMC data for various coupling strengths
$m_0$. Interestingly, the equilibrated model describes the exact data very accurately from weak up to moderate
electron-phonon coupling $m_0\approx 1$, while deviations appear for stronger couplings $m_0 \gtrsim 1$
and voltages beyond the first plateau $e V>2\hbar\omega_0$. For $m_0>1$ nonequilibrium effects are stronger
and the corresponding master equation (\ref{NEQMaster1}) gives a better description of higher order resonant
steps. Moreover, as already addressed above, even in this low temperature domain the approximate description
provides quantitatively reliable results.
In Fig.~\ref{MC3} the frequency of the phonon mode is fixed and only the electron-phonon coupling is tuned
over a wider range. For strong coupling (here $m_0=2$) the equilibrated (nonequilibrated) model predicts a
smaller (larger) current than the exact one in contrast to the situation for smaller $m_0$. This phenomenon
directly results from what has been said above in Sec.~\ref{sec3}: for stronger coupling the Franck-Condon
overlaps favor also higher lying phonon states that are suppressed by a thermal distribution.
After all, the approximate models give not only qualitatively a correct picture of the exact $IV$ curves,
but provide even quantitatively a reasonable description in this low temperature domain.

\begin{figure}
\begin{center}
\includegraphics[width=8.5cm]{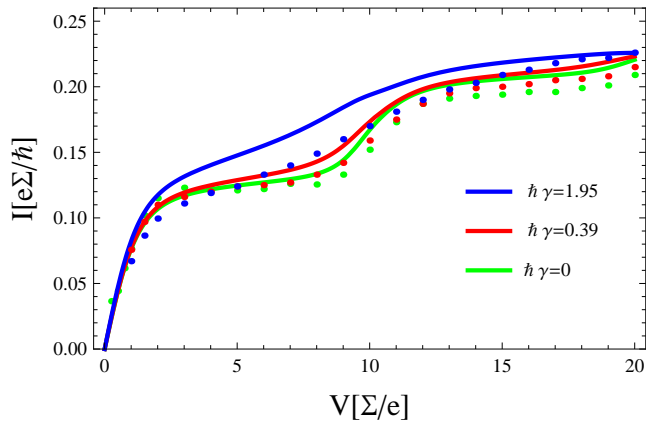}
\end{center}
\caption{$IV$-characteristics in presence of a secondary heat bath interacting with the phonon with various
coupling constants $\gamma$.
Shown are approximate results (solid) using (\ref{P(E)final}) and diagMC data (dots); energies are scaled
with $\Sigma$. Other parameters are $k_{\rm B} T/\Sigma=0.2$, $m_0=4/5$, $\sigma=0.2$.}
\label{MC1}
\end{figure}

In a next step the coupling to  a secondary bath is turned on ($\gamma\neq 0$) enforcing equilibration of the
phonon mode, see (\ref{P(E)final}). The expectation is that in this case departures from the equilibrated model
are reduced. In Fig.~\ref{MC1} data are shown for a ratio $m_0=4/5$ where deviations occur at larger voltages
as observed in the previous figures. Obviously, due to the damping of the phonon mode the resonant steps are
smeared out with increasing $\gamma$. However, the approximate model predicts this effect to be more pronounced
as compared to the exact data, particularly for stronger coupling $\hbar\gamma/\Sigma>1$, while still
$\gamma/\omega_0<1$.
In fact, in the limit of very large coupling only the $k=l=0$ contribution to (\ref{P(E)final}) survives so
that at zero temperature one arrives at
\begin{equation}
\lim_{\gamma\to\infty}I(V)=I_\infty \frac{2}{\pi} {\rm arctan}\left(\frac{e V}{\hbar\Gamma_{\rm tot, 0}(V)}\right)
\end{equation}
with the current at large voltages $I_\infty=e\Sigma/4\hbar$ and $\Gamma_{\rm tot, 0}(V)\leq \Sigma/\hbar$
where equality is approached for $V\to \infty$.
It seems that a broadened equilibrium distribution of the phonon induced by the secondary bath according to
(\ref{P(E)final}) overestimates the broadening of individual levels. Since the approach is exact in the limit
$m_0\to 0$, the deviations appearing in Fig.~\ref{MC1} are due to intimate electron-phonon-secondary bath
correlations not captured by the rate approach. In the overdamped regime, i.e.\ $\gamma/\omega_0>1$, the
dynamics of the phonon mode slows down and may become almost static on the time scale of the charge transfer.
In this adiabatic regime an extended version of the master equation (\ref{NEQMaster1}) is not trivial since
the conventional eigenstate representation becomes meaningless. One should then better switch to phase-space
coordinates and develop a formulation based on a Fokker-Planck or Smoluchowski equation for the phonon.
This will be the subject of future research.

\begin{figure}
\begin{center}
\includegraphics[width=8.5cm]{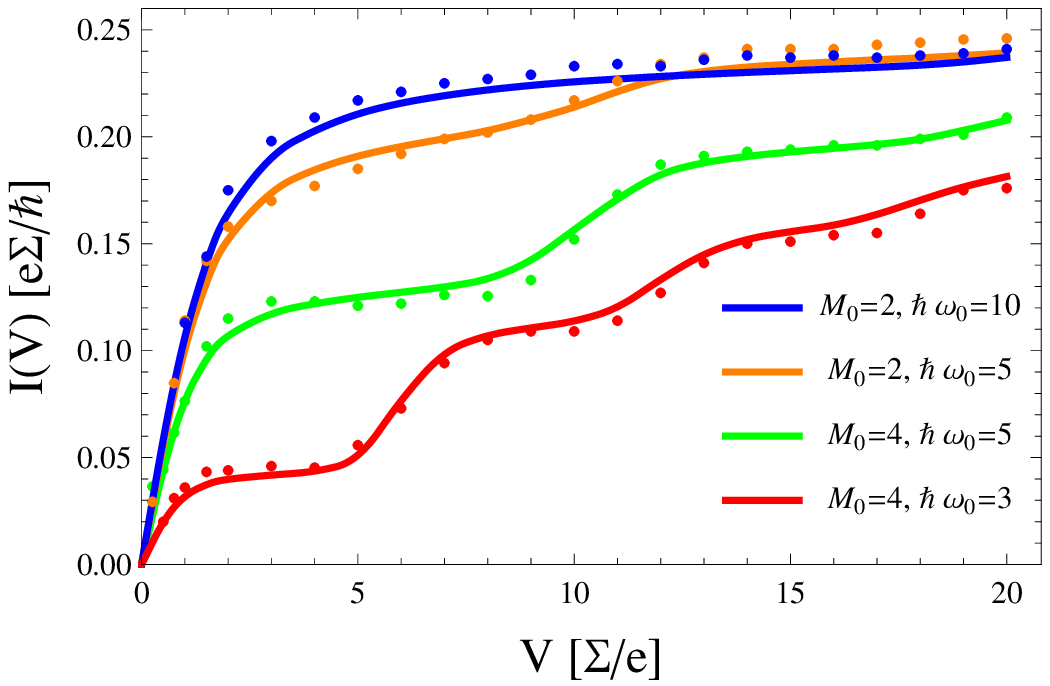}
\end{center}
\caption{Same as in Fig.~\ref{MC2}, but for nonequilibrated phonons based on an extended master equation
(solid) in comparison to exact diagMC data (dots).}
\label{MC4}
\end{figure}
\begin{figure}
\begin{center}
\vspace*{3mm}
\includegraphics[width=8.5cm]{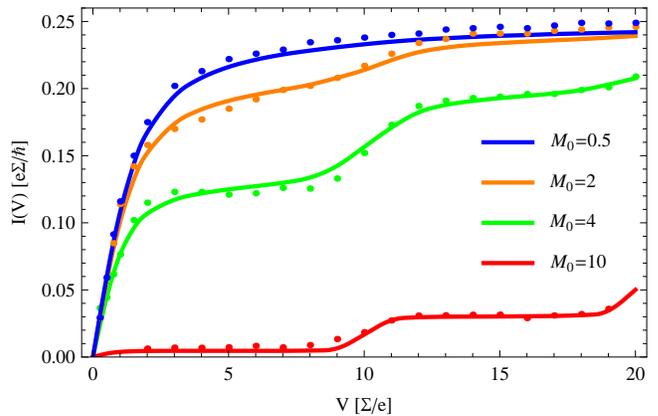}
\end{center}
\caption{Same as in Fig.~\ref{MC3}, but for nonequilibrated phonons based on an extended master equation
(solid) in comparison to exact diagMC data (dots).}
\label{MC5}
\end{figure}
The essence of this comparison is that, as anticipated from physical arguments already in Sec.~\ref{sec1},
a rate description does indeed provide even quantitatively accurate results in the regime of weak to moderate
electron-phonon coupling $m_0<1$ and for all $\sigma/m_0$.
Deviations that occur for larger values of $m_0$ can partially be explained by  nonequilibrium distributions
in the phonon distribution, where, however, the master equation approach seems to overestimate this effect.
In order to obtain some insight into the nature of this deficiency, a minimal approach consists of adding to
(\ref{NEQMaster1}) with the extension (\ref{replacement}) a mechanism that enforces relaxation to thermal
equilibrium with a single rate constant $\Gamma_0$ that serves as a fitting parameter. Accordingly, the
respective time evolution equation for $P_q^n(t)$ receives an additional term
$-\Gamma_0 [P_q^n(t)-P_\beta^n/2]$ with the Boltzmann distribution for the bare phonon degree of freedom
$P_\beta^n$. Corresponding results for the same parameter range as in Figs.~\ref{MC2}, \ref{MC3} are
shown in Figs.~\ref{MC4}, \ref{MC5} in comparison with exact diagMC data. There, the {\em same} equilibration
rate $\hbar\Gamma_0/\Sigma=0.25$ is used for {\em all} parameter sets. Astonishingly, this procedure provides
an excellent agreement over the full voltage range. It improves results particularly in the range of moderate
to stronger electron phonon coupling, but has only minor impact for $m_0<1$. The indication is thus that
electron-phonon correlations neglected in the original form of the master equation have effectively the
tendency to support faster thermalization of the phonon. Indeed, preliminary results with a generalized master
equation where the coupling between diagonal (populations) and off-diagonal (coherences) elements of the
reduced charge-phonon density matrix are retained  (no RWA approximation) indicate that this coupling leads
to an enhanced phonon-lead interaction and thus to enhanced phonon equilibration.
\begin{acknowledgments}
We thank L. M\"uhlbacher for fruitful discussions and for providing numerical data of Ref.~\cite{Lothar}.
Financial support was provided by the SFB569, the Baden-W\"urttemberg Stiftung, and the German-Israeli Foundation (GIF).
\end{acknowledgments}

\end{document}